\documentclass[sn-vancouver,Numbered]{sn-jnl}

\usepackage{graphicx}%
\usepackage{multirow}%
\usepackage{amsmath,amssymb,amsfonts}%
\usepackage{amsthm}%
\usepackage{mathrsfs}%
\usepackage[title]{appendix}%
\usepackage{xcolor}%
\usepackage{textcomp}%
\usepackage{manyfoot}%
\usepackage{booktabs}%
\usepackage{algorithm}%
\usepackage{algorithmicx}%
\usepackage{algpseudocode}%
\usepackage{listings}%
\usepackage{graphicx}
\usepackage{url}

\usepackage{amsfonts}
\usepackage{amssymb}
\usepackage{bm}
\renewcommand{\arraystretch}{1.2}
\usepackage{eqnarray,amsmath}
\usepackage{physics}
\usepackage{fancybox}
\usepackage{subcaption,float}
\usepackage{caption}
\usepackage{float}
\usepackage{tikz-cd}
\usepackage{pgf,tikz}
\usepackage{tikz}
\usetikzlibrary{calc}
\usetikzlibrary{intersections}
\usetikzlibrary{shapes.arrows}
\usetikzlibrary{arrows.meta}
\usetikzlibrary{shapes}
\usetikzlibrary{patterns}
\usepackage{pgfplots}
\usepackage{graphicx}
\usetikzlibrary{decorations.markings}
\usetikzlibrary{arrows}

\usepackage[autostyle]{csquotes}
\usepackage{bm}
\usepackage{array,mathtools}
\usepackage{textcomp}

\theoremstyle{thmstyleone}%

\theoremstyle{thmstyletwo}%

\theoremstyle{thmstylethree}%
\usepackage{mathtools}
\begin{document}
	
	\title[Article Title]{Design of two-dimensional reflective imaging systems: An approach based on inverse methods}

	\author*[1]{\fnm{Sanjana} \sur{Verma}}\email{s.verma@tue.nl}
	
	\author[1]{\fnm{Martijn} \sur{J.H. Anthonissen}}\email{m.j.h.anthonissen@tue.nl}
	\equalcont{These authors contributed equally to this work.}
	
	\author[1]{\fnm{Jan} \sur{H.M. ten Thije Boonkkamp}}\email{j.h.m.tenthijeboonkkamp@tue.nl}\equalcont{These authors contributed equally to this work.}
	
		\author[2,1]{\fnm{Wilbert} \sur{L. IJzerman}}\email{wilbert.ijzerman@signify.com}
	\equalcont{These authors contributed equally to this work.}

	\affil[1]{\orgdiv{Department of Mathematics \& Computer Science}, \orgname{Eindhoven University of Technology}, \orgaddress{\street{PO Box 513}, \postcode{5600 MB} \city{Eindhoven}, \country{The Netherlands}}}
	
	\affil[2]{\orgname{Signify Research}, \orgaddress{\street{High Tech Campus 7},  \postcode{5656 AE} \city{Eindhoven},\\ \country{The Netherlands}}}

	
	\abstract{Imaging systems are inherently prone to aberrations. We present an optimization method to design two-dimensional (2D) freeform reflectors that minimize aberrations for various parallel ray beams incident on the optical system. We iteratively design reflectors using inverse methods from non-imaging optics and optimize them to obtain a system that produces minimal aberrations. This is done by minimizing a merit function that quantifies aberrations and is dependent on the  energy distributions at the source and target of an optical system, which are input parameters essential for inverse freeform design. The proposed method is tested for two configurations: a single-reflector system and a double-reflector system. Classical designs consisting of aspheric elements are well-known for their ability to minimize aberrations. We compare the performance of our freeform optical elements with classical designs. The optimized freeform designs outperform the classical designs in both configurations.}


	\maketitle
	
	\section{Introduction}\label{sec1}

    Recent technological advancements have lead to an increasing demand for high-quality imaging optical systems. They are crucial for diverse applications like metrology in the semiconductor industry, imaging and diagnosis in medical science and astronomical observations. However, imaging systems are adversely affected by \textit{aberrations}, i.e., small deviations from an ideal image \cite{hecht,korsch,braat}. It is essential to analyze aberrations and develop methods to design precise imaging systems.

    Aberration theory (AT) \cite{hecht,korsch} is used to investigate different types of aberrations by expressing the optical map as a function of source coordinates and parameters that describe the optical elements. It has been utilized to find conditions on the optical system layout and shapes of the optical elements for aberration correction. Traditionally, a systematic design of imaging systems was based on the mathematical analysis of ray propagation through an optical system \cite{braat}.  
    
    Conventional designs comprising of aspheric shapes encounter challenges in effectively correcting all aberrations, leading to compromises in overall image quality. This has catalyzed a growing interest in  freeform surfaces, i.e., surfaces without any symmetry. These surfaces provide increased degrees of freedom to optimize optical systems for perfect imaging. 
	
	 Imaging systems consisting of freeform elements are computed and analyzed using two main approaches, numerical and analytical. The Simultaneous Multiple Surfaces (SMS) method \cite[Chap.~9]{chaves} is a numerical method that was originally used to compute multiple surfaces simultaneously for non-imaging systems by employing geometrical optics (GO). It has been extended to calculate designs for imaging applications by imposing conditions to eliminate certain aberrations for a particular set of rays \cite{SMS,SMS2}. In \cite{fabian}, Duerr et al. presented a `first time right' design method based on Fermat's principle. With this method, a highly systematic generation and evaluation of imaging designs is possible. Designs are typically analyzed by employing a commercial software like Zemax to calculate aberrations using raytracing \cite{braat,SMS,SMS2,fabian,julie,carmela}. 
	
    Although numerical methods are effective design methods, analytic methods are necessary for an in-depth analysis about the causes of aberrations.	An analytical approach based on a Zernike polynomial  expansion of optical surfaces combined with Nodal Aberration Theory is used to determine aberrations \cite{NAT,NAT2}. It provides insights about the contributions of each optical surface, thereby enabling an optical designer to improve designs. 
	
	Classical methods aimed at optimizing imaging systems minimize a merit function that quantifies image quality \cite[Chap.~6]{braat}. The merit function is parameterized by various design specifications of the optical system elements like deformation constants, radii of curvature, location, and additional coefficients for defining freeform shapes as perturbations to conic shapes. The merit function is usually quite complex and non-linear in nature. Multi-parametric optimization techniques are employed for determining the minimum and quite often the function lands in a narrow landscape of a local minimum. It is very unlikely that a design form obtained after optimization corresponds to the global minimum \cite{braat,fabian,SMS}. Thus, it is imperative to start from a good initial guess. Determining a good initial guess can be time consuming and challenging as it requires a lot of expertise.

	 AT is limited to calculating aberrations for aspheric optical elements and applies to a relatively small class of optical systems. Moreover, calculating analytical expressions for highly sophisticated systems with multiple elements and freeform shapes can be very laborious. The SMS and `first time right' design methods provide excellent starting points for optimization. However, all the above mentioned methods for computing and optimizing freeform imaging elements are forward methods. Our focus is to develop methods for designing imaging systems consisting of fully freeform elements using inverse methods.
	
	\textit{Inverse methods} in non-imaging optics are used to compute freeform surfaces that convert a given source distribution into a desired target distribution. Optical systems are modeled by nonlinear partial differential equations of Monge-Amp\`{e}re type \cite{ecmi1}. These are obtained by employing optimal transport theory and combining the principles of GO with conservation of energy. These methods can also be utilized to design fully \textit{freeform imaging} systems if aberrations are quantified using a merit function dependent on the parameters of inverse design. In this paper, we formulate a new method for optimizing two-dimensional imaging systems using non-imaging design methodologies. Furthermore, we compare our optimized optical systems with classical designs \cite{korsch} that correct aberrations and have been obtained using the principles of AT.
	
	The contents of this contribution are the following. The theory for quantifying aberrations is introduced in Sect.~\ref{sec: framework}. It also outlines the construction of a merit function for optimization and its dependence on the inverse design parameters, i.e., energy distributions at the source and the target of an optical system. Inverse methods to compute freeform reflectors are presented in Sect.~\ref{sec: inverse: one reflector}-\ref{sec: inverse: two reflectors}. An explanation of the optimization algorithm using the \textit{Nelder-Mead simplex method} \cite[p.~502-507]{ecmi3} can be found in Sect.~\ref{sec:algorithm}. Sect.~\ref{sec: results} contains numerical results for two optical system configurations, a single-reflector system with a parallel source and a near-field target, and a double-reflector system with a parallel source and a point target. A comparison between the performance of the freeform reflectors obtained after optimization and classical design forms is shown. Finally, some concluding remarks and the scope for extensions are presented in Sect.~\ref{sec: conclusions}.
	
	\section{Methodological framework} \label{sec: framework}
	In this section, we present the method of quantifying aberrations and a brief overview of our approach for inverse design of freeform reflectors suitable for perfect imaging.
	\subsection{Quantifying aberrations} \label{sec: abb}
	Ray aberrations are evaluated using the root-mean-square (RMS) spot size that represents the geometric blur \cite{julie}. For 2D optical systems, we denote the source and target coordinates of any ray by $x$ and $y$, respectively. A ray tracer determines the path of light rays through an optical system. Thus, we can easily find the target coordinate $y$ for any ray starting from the source coordinate $x$. During ray tracing, rays are generated from the source randomly, resulting in random target coordinates $y$. Therefore, $y$ is a stochastic variable. The RMS spot size can be seen as an indicator of the extent of spread or variability of target coordinates. Just as the standard deviation describes how data points deviate from the statistical mean, the RMS spot size measures the average distance of target coordinates from an ideal non-aberrated spot. This comparison allows us to view the RMS spot size as the standard deviation for target coordinates. 
	
	An optical system consists of on-axis and off-axis light rays, depending on the angle they make with the optical axis, i.e., the $z$-axis. Let $\alpha$ be the angle, measured counterclockwise from the $z$-axis to the incoming light rays. We denote the RMS spot size corresponding to a parallel ray beam making an angle $\alpha$ with the optical axis by $W_{\alpha}$,
\begin{equation}	
	\label{eq: RMS}
	W_{\alpha}=\sqrt{\rm{Var}[\rm{y}]},
	\end{equation}
where $\rm{Var}[\rm{y}]$ is the variance of target coordinates $y$.

\begin{figure}[H]
			\vspace{-3mm}
		\centering
		\begin{subfigure}[b]{0.5\textwidth}		
			\centering
			\includegraphics[trim={0cm 0cm 0.8cm 1.85cm},clip,scale=0.9]{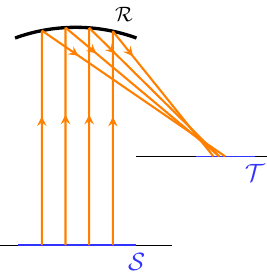}
			\caption{Parallel-to-near-field.}
		\end{subfigure}%
		\hspace{-1cm} 
		\begin{subfigure}[b]{0.5\textwidth}
			\centering
			\includegraphics[trim={0.3cm 0cm 0.3cm 1.5cm},clip,scale=0.75]{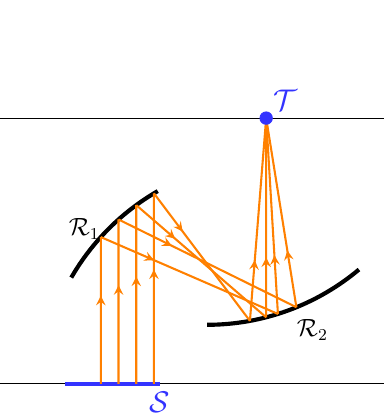}
			\caption{Parallel-to-point.}
		\end{subfigure}
		\caption{Optical systems with on-axis parallel light rays.}
		\label{fig: both systems}
		\vspace{-3mm}
	\end{figure}

	We refer to	the case $\alpha=0^{\circ}$ as the \textit{base case}, and it denotes on-axis light rays. In this paper, we will consider two configurations, a single-reflector parallel-to-near-field system, and a double-reflector parallel-to-point system (see Fig.~\ref{fig: both systems}).
	\subsection{Overview of the method} \label{sec: overview}
	In non-imaging optics, inverse methods aim to compute freeform optical elements that convert a given source distribution $f$ to a desired target distribution $g$ \cite{ecmi1}. For 2D systems, the shape of a reflector is the solution of an ordinary differential equation (ODE), which is determined by combining the optical map and conservation of energy. We use this inverse freeform reflector design methodology to optimize imaging designs. In this paper, we focus only on symmetric systems because many aberrations are eliminated in such systems due to symmetry of the optical components \cite{korsch}. 
		
	In Sect.~\ref{sec: inverse: one reflector}-\ref{sec: inverse: two reflectors}, we present freeform inverse design methods  for the base case configurations. We show that the shape of the reflectors is dependent on the ratio $\Upsilon$ of the energy distributions at  the source and target. Furthermore, we prove that if this ratio is an even function, the reflectors designed will be symmetric. 

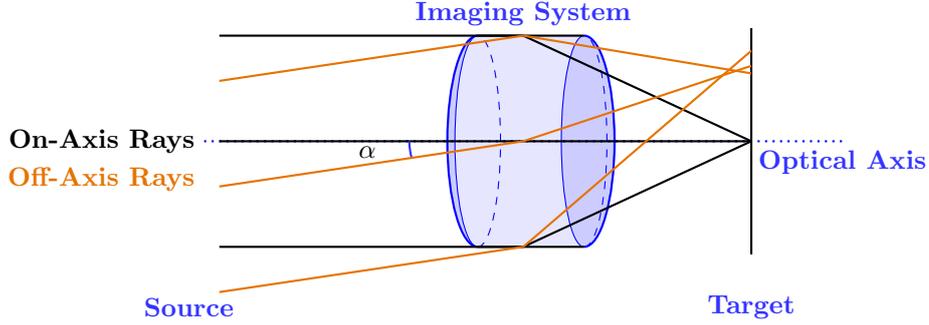
\begin{figure}[H]
				\vspace{-3mm}
			\centering
			\begin{tikzpicture}
				\def\z{-0.6}; 
			\fill [blue!20] (\z,1.4) arc (90:270:0.4cm and 1.4cm);
			\fill [blue!10] (\z,0) ellipse (0.3cm and 1.4cm);
			\fill [blue!20] (\z+1.4,-1.4) arc (-90:90:0.4cm and 1.4cm);
			\fill [blue!10] (\z,1.4) -- (\z+1.4,1.4) -- (\z+1.4,-1.4) -- (\z,-1.4) -- cycle;
			\fill [blue!20] (\z+1.4,0) ellipse (0.3cm and 1.4cm);
			
			\draw [blue, dashed] (\z,-1.4) arc (-90:90:0.3cm and 1.4cm);
			\draw [blue] (\z,1.4) arc (90:270:0.3cm and 1.4cm);
			
			\draw [thick,blue] (\z,1.4) arc (90:270:0.4cm and 1.4cm);
			
			\draw [blue] (\z+1.4,1.4) arc (90:270:0.3cm and 1.4cm);
			\draw [blue, dashed] (\z+1.4,-1.4) arc (-90:90:0.3cm and 1.4cm);
			
			\draw [thick,blue] (\z+1.4,-1.4) arc (-90:90:0.4cm and 1.4cm);
			
			\draw [thick] (\z,1.4) -- +(1.4,0);
			\draw [thick] (\z,-1.4) -- +(1.4,0);
			
			
				\node at (0, 1.7) {\bf{\textcolor{blue!80}{Imaging System}}};
				\node at (-4.4,-2.2) {\bf{\textcolor{blue!80}{Source}}};
				\node at (3,-2.2) {\bf{\textcolor{blue!80}{Target}}};
				\node[left] at (-4.2,0) {\bf{On-Axis Rays}};
				
				\node[below] at (4.2,0) {\bf{\textcolor{blue!80}{Optical Axis}}};    
				\draw[thick,dotted, blue] (-4.2,0) -- (4.2,0);  
				\draw[thick] (-4,1.4) -- (0,1.4);
				\draw[thick] (-4,0) -- (0,0);
				\draw[thick] (-4,-1.4) -- (0,-1.4);
				\draw[thick] (0,1.4) -- (3,0);
				\draw[thick] (0,0) -- (3,0);
				\draw[thick] (0,-1.4) -- (3,0);
				\draw[thick] (3,-1.5) -- (3,1.5);

				\node[left] at (-4.2,-0.5) {{\bf{\textcolor{orange!90!black}{Off-Axis Rays}}}};
				\draw[orange!90!black, thick] (-4,0.8) -- (0,1.4);
				\draw[orange!90!black, thick] (-4,-0.6) -- (0,0);
				\draw[orange!90!black, thick] (-4,-2) -- (0,-1.4);
				\draw[orange!90!black, thick] (0,1.4) -- (3,0.9);
				\draw[orange!90!black, thick] (0,0) -- (3,1);
				\draw[orange!90!black, thick] (0,-1.4) -- (3,1.2);
				
				\draw[blue!80, thick] (-1.5,0) arc (-180:-165:0.8); 
				\node at (-2.05,-0.15) {$\bf{\alpha}$}; 
				
			\end{tikzpicture}
			\caption{Off-axis rays form a blur in the target.}
			\label{fig: blur}
				\vspace{-3mm}
		\end{figure}

 The base case configurations enable us to compute reflectors that form a perfect image for on-axis parallel rays. When a set of off-axis rays passes through an optical system, the resulting image is an aberrated spot. This spot differs from an ideal, in-focus spot produced by on-axis rays (see Fig.~\ref{fig: blur}).  The RMS spot size for any parallel beam passing through an optical system depends on the angle of incoming rays and the energy distributions used to compute reflectors. 

Our goal is to design reflectors that minimize aberrations for off-axis ray beams. In order to optimize an imaging system, we first construct a merit function that can measure the aberrations of various ray beams (on-axis and off-axis) passing through the system. The value of the merit function is the RMS of the spot sizes for the parallel ray beams inclined at an angle $\alpha$ with respect to the optical axis. This value is unique for given energy distributions and incorporates a range of angles taken into consideration. Consider a non-uniform source distribution $f$ and a uniform target distribution $g$, which is chosen such that the global energy balance at the source and target is satisfied. The shape of the reflectors is then dependent on $f$ only. Thus, we want to minimize an objective function dependent on the energy distribution $f$, and design an optical system that minimizes aberrations for various light beams. We use the inverse freeform reflector design methodology to compute freeform reflectors and subsequently use optimization methods to determine designs that minimize aberrations.

\section{Inverse freeform design: parallel-to-near-field one-reflector system} \label{sec: inverse: one reflector}

\subsection{{Optical map}} 
\begin{figure}[H]
	\vspace{-2cm}
	\centering
	\includegraphics[width=0.55\textwidth]{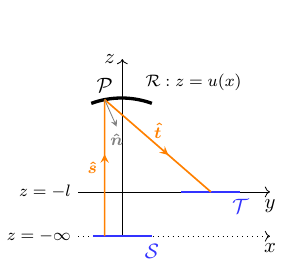}
	\caption{Parallel source to near-field target with on-axis parallel rays.}
	\label{fig:one reflector}
	\vspace{-5mm}
\end{figure}

Consider the 2D optical system with the $z$-axis as the optical axis as shown in Fig.~\ref{fig:one reflector}. The source $\mathcal{S}$ and target $\mathcal{T}$ are parametrized with the $x$- and $y$-coordinates, respectively. A parallel source forms a near-field image in $z=-l$ after hitting a reflector $z=u(x)$. The law of reflection is employed to find the optical mapping $y=\widetilde{m}(x,u,u'),$ which connects the coordinates of the source and target domains. Consider a ray emitted from the source, which has a unit direction vector\linebreak ${\boldsymbol{\mathit{\hat{s}}}}=(0,1)^T$. The downward normal vector to the reflector is
\begin{equation}
	\label{eq:normal}
	\mathit{\boldsymbol{{n}}}=\begin{pmatrix}u'\\-1\end{pmatrix}, \quad {\boldsymbol{\hat{\mathit{n}}}}=\frac{\mathit{\boldsymbol{{n}}}}{\abs{\mathit{\boldsymbol{{n}}}}}.
\end{equation}
The reflected direction ${\boldsymbol{\mathit{\hat{t}}}}$ is determined by applying the vectorial law of reflection \linebreak\cite[p.~22-24]{ecmi1}, ${\boldsymbol{\mathit{\hat{t}}}}={\boldsymbol{\mathit{\hat{s}}}}-2({\boldsymbol{\mathit{\hat{s}}}}\bm{\cdot}{\boldsymbol{\mathit{\hat{n}}}}){\boldsymbol{\mathit{\hat{n}}}}$, resulting in 
\begin{equation}
	\label{eq:reflected direction}
	{\boldsymbol{\mathit{\hat{t}}}}=\frac{1}{(u')^2+1}\begin{pmatrix}2u'\\(u')^2-1\end{pmatrix}.
\end{equation}
	To compute the optical mapping, we determine the point of intersection of the reflected ray with the target $z=-l$ by eliminating $\lambda$ from the system
	\begin{subequations}
		\begin{align}
			\label{eq:1}
			y&=x+\lambda t_1,\\
			\label{eq:2}
			-l&=u+\lambda t_2.
		\end{align}
		\label{eq:system of equation}
	\end{subequations}
	Substituting $\boldsymbol{\mathit{\hat{t}}}=(t_1,t_2)^T$ from Eq.~(\ref{eq:reflected direction}) in Eq.~(\ref{eq:system of equation}) gives us the optical map
	\begin{equation}
		\label{eq:opticalmap}
		y=\widetilde{m}(x,u,u')=x-\frac{2u'(u+l)}{(u')^2-1}.
	\end{equation}
\noindent	\subsection{{Energy conservation}}
	The emittance at the source $\mathcal{S}$ and the illuminance at the target $\mathcal{T}$ is given by $f(x)$ and $g(y),$ respectively. The law of conservation of energy gives us 
	\begin{equation}
		\label{eq:energy balance 1}
		\int_{\mathcal{A}}f(x)\,\text{d}x=\int_{m(\mathcal{A})}g(y)\,\text{d}y,
	\end{equation}
	for an arbitrary subset $\mathcal{A}\subseteq\mathcal{S}$ and image set $m(\mathcal{A})\subseteq\mathcal{T}$. We assume that there exists an optical map $y=m(x)$ such that the total energy at the source and the target is conserved. The energy distributions $f(x)$ and $g(y)$ have to satisfy the global energy balance, implying that Eq.~(\ref{eq:energy balance 1}) holds for $\mathcal{A}=\mathcal{S}$ and $m(\mathcal{A})=\mathcal{T}$. Substituting $y=m(x)$ in Eq.~(\ref{eq:energy balance 1}) gives us
	\begin{equation}
		{m'(x)}=\pm\frac{f(x)}{g(m(x))},
		\label{eq: mapping ode1}
	\end{equation}
	subject to the transport boundary condition $m(\partial \mathcal{S})=\partial \mathcal{T}$. This condition is imposed as a consequence of the edge ray principle \cite{EDGE} and implies that all light from the source arrives at the target. Due to global energy conservation, solving Eq.~(\ref{eq: mapping ode1}) as an initial value problem (IVP) ensures that the boundary condition at the opposite endpoint is also satisfied.
	
\noindent	\subsection{{Freeform reflector}}
	We solve Eq.~(\ref{eq: mapping ode1}) (with either $+$ or $-$ sign) and obtain a numerical solution for the mapping. We \textit{interpolate} the numerical solution to obtain the function $y=m(x)$. This mapping should be equivalent to the optical map $y=\widetilde{m}(x,u,u')$ as given in Eq.~(\ref{eq:opticalmap}). Thus, we substitute $y=m(x)$ in Eq.~(\ref{eq:opticalmap}) and solve for $u'$ to obtain the differential equations
	\begin{equation}
			\label{eq: reflector ode}
		u'(x)=\frac{u(x)+l}{m(x)-x}\left(-1\pm\sqrt{1+\left(\frac{m(x)-x}{u(x)+l}\right)^2}\right), 
	\end{equation}
	subject to the condition $u(0)=0$, that gives the location of the vertex of the reflector at the origin of the optical system. This choice makes it easier to analyze aberrations analytically. The reflector $z=u(x)$ is obtained by numerically solving Eq.~(\ref{eq: reflector ode}). We can only choose the solution corresponding to the ODE with the positive sign, since Eq.~(\ref{eq: reflector ode}) is not defined at $x=0$ otherwise.
	
	 \subsection{{Condition for symmetry}}
	We introduce the notation $\Upsilon(x)=f(x)/g(m(x))$. Suppose that $\Upsilon(x)$ is an even function, i.e.,
		$\Upsilon(x)=\Upsilon(-x)$. We prove that this implies $u(x)=u(-x)$, i.e., $u(x)$ is even.
	 Let us assume we have a symmetric source domain $x\in[-a,a]$, for some $a>0$. For an even function $\Upsilon(x)$, Eq.~(\ref{eq: mapping ode1}) leads to $m'(x)=m'(-x)$, which after integration gives $m(x)=-m(-x)+C$,
where $C$ is a constant of integration. We want to design a symmetric reflector, which requires $u'(0)=0$. Using the law of reflection, it is trivial to show that $u'(0)=0$ implies $m(0)=0$, and subsequently $C=0$. 

Eq.~(\ref{eq: reflector ode}) gives an IVP for the shape of the freeform reflector
\begin{subequations}
			\begin{align}		
	u'(x)&=h(x,u), \quad u(0)=0,\quad x\in[-a,a],\\
		\intertext{where,}  
		h(x,u)&=\frac{u(x)+l}{m(x)-x}\left(-1+\sqrt{1+\left(\frac{m(x)-x}{u(x)+l}\right)^2}\right).		
		\end{align}
			\label{eq: freeform reflector ode}
\end{subequations}
We assume that $h(x,u)$ is defined for $(x,u)\in D=[-a,a]\cross[-b,b]$, for some $a,b>0$. We prove that Eq.~(\ref{eq: freeform reflector ode}) has a unique solution in $D$.  
The function $h(x,u)$ is defined and continuous for all $x$, except for $x=0$. It is evident from Eq.~(\ref{eq:opticalmap}) that $\abs{(m(x)-x)/(u(x)+l)}<1$ for $u'(x)\in(1-\sqrt{2},-1+\sqrt{2})$. Thus, $h(x,u)$ can be expressed as a binomial series. We now check continuity of the function at $x=0$. We determine
\begin{equation}
\displaystyle{\lim_{x\rightarrow0}} h(x,u)
	=\displaystyle{\lim_{x\rightarrow0}}\left(\frac{1}{2}\left(\frac{m(x)-x}{u(x)+l}\right)-\frac{1}{8}\left(\frac{m(x)-x}{u(x)+l}\right)^3+\cdots\right)=0.
	\end{equation}
The limit of $h(x,u)$ is equal to the value of the function at $x=0$. Therefore, $h(x,u)$ is continuous in $D$. The partial derivative of $h(x,u)$ with respect to $u$ is given by
\begin{equation}
	\frac{\partial h}{\partial u}(x,u)=
		\frac{1}{m(x)-x}\left(-1+\left(\sqrt{1+\left(\frac{m(x)-x}{u(x)+l}\right)^2}\right)^{-1}\right).
\end{equation}
We express $\partial h/\partial u$ as a binomial series and observe that it is continuous in $D$. Since $D$ is a compact domain, $\partial h/\partial u$ is a bounded function in $D$ \cite[p.~130]{boundedness}. Using the Existence and Uniqueness Theorem for ODEs \cite[p.~112]{ode-uniqueness}, Eq.~(\ref{eq: freeform reflector ode}) has a unique solution $u(x)$.

Let us define a function $v(x)=u(-x)$. Since $m(x)$ is an odd function, (\ref{eq: freeform reflector ode}) implies 
\begin{equation}
	\begin{aligned}
		v'(x)&=-u'(-x)\\
		&=-\frac{u(-x)+l}{m(-x)+x}\left(-1+\sqrt{1+\left(\frac{m(-x)+x}{u(-x)+l}\right)^2}\right)\\
		&=\frac{v(x)+l}{m(x)-x}\left(-1+\sqrt{1+\left(\frac{m(x)-x}{v(x)+l}\right)^2}\right).
	\end{aligned}
\label{eq: v(x)}
\end{equation}
Eq.~(\ref{eq: v(x)}) has the solution $v(x)$ and is equivalent to Eq.~(\ref{eq: freeform reflector ode}). We have already shown that Eq.~(\ref{eq: freeform reflector ode}) has a unique solution. This results in 
\begin{equation}
u(x)=v(x)=u(-x).
\end{equation}
Therefore, the freeform reflector is symmetric about the $z$-axis.

\section{Inverse freeform design: Parallel-to-point double-reflector system} \label{sec: inverse: two reflectors}
	
	\subsection{{Optimal transport formulation}}
	  In optimal mass transport problems, we find a transfer plan which relocates all mass from one location to another while minimizing the total transportation cost. For designing freeform reflectors, we consider our problem analogous to an optimal transport problem. The reflectors are designed to transfer energy from a source to a target with a minimum cost function. They are designed in a manner that a unique optical map based on the law of reflection  conserves energy throughout the system. We want to obtain the following relation for our optical system
	  \begin{equation}
	  	\label{eq: cost}
	  	u_1(x)+u_2(y)=c(x,y),
	  \end{equation}
	  where $c(x,y)$ is the cost function in the framework of optimal mass transport problems \cite[p.~57,~65-68]{nitin}, and $u_1(x)$ and $u_2(y)$ are related to the locations of the freeform surfaces.
	  
	   \begin{figure}[H]
	  	\vspace{-5mm}
	  	\centering
	  	\includegraphics[width=0.75\textwidth]{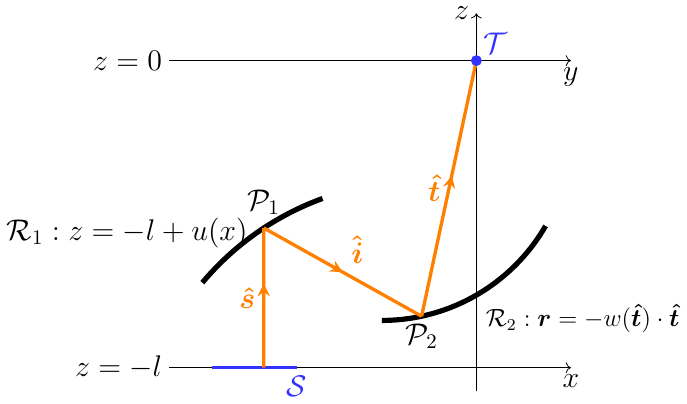}
	  	\caption{Parallel source to point target with on-axis parallel rays.}
	  	\label{fig:two reflectors}
	  	\vspace{-3mm}
	  \end{figure}
	   Consider a 2D optical system consisting of two reflectors $\mathcal{R}_1,$ $\mathcal{R}_2$ and a set of parallel rays forming a point image in the target plane as shown in Fig.~\ref{fig:two reflectors}. A parallel source in $z=-l$ emits rays with unit direction vector ${\boldsymbol{\mathit{\hat{s}}}}=(0,1)^T$. We choose the point target to be located at the origin $\mathcal{O}$ of the target coordinate system. The rays hit the target plane with unit direction vectors ${\boldsymbol{\mathit{\hat{t}}}}=(t_1,t_2)^T.$
	  The first reflector, $\mathcal{R}_1$: $z=-l+u(x)$, is defined by the perpendicular distance from the source plane $z=-l.$ The second reflector, $\mathcal{R}_2$: $\boldsymbol{\mathit{\hat{r}}}=-w(\boldsymbol{\mathit{\hat{t}}})\boldsymbol{\mathit{\hat{t}}}$, is defined by the radial distance from the target $w=w(\boldsymbol{\mathit{\hat{t}}})$.  We denote with $\mathcal{P}_1\left(x,-l+u(x)\right)$ and $\mathcal{P}_2\left(-w({\boldsymbol{\mathit{\hat{t}}}})t_1,-w({\boldsymbol{\mathit{\hat{t}}}})t_2\right)$, the points where a ray hits the first and second reflector, respectively. The distance between those two points is denoted by $d$. The optical path length, $L$, is given by
	  \begin{equation}
	  	\label{eq:L}
	  	L=u(x)+d+w({\boldsymbol{\mathit{\hat{t}}}}),
	  \end{equation}
	 and is constant as a consequence of the Malus-Dupin theorem \cite[p.~130]{M-Dthm}. 
	  	 
	 We introduce the following notation for an arbitrary ray in the optical system. The positions of the ray at the source and target are given by $q_\text{s}=x$ and $q_\text{t}=0,$ respectively. The projections of the unit direction vectors on the source and target planes are $p_\text{s}=0$ and $p_\text{t},$ respectively. In terms of Hamiltonian characteristics, the optical path length (OPL), denoted by $L,$ is equal to the point characteristic $V$. We can write
	\begin{equation}
		V(q_\text{s},q_\text{t})=u(x)+d+w({\boldsymbol{\mathit{\hat{t}}}}).
	\end{equation}
 Since, $V$ is dependent on the position of the source and  the direction at the target, we work with the mixed characteristic of the first kind $W=W(q_\text{s},p_\text{t})$, given by
	\begin{equation}
		W(q_\text{s},p_\text{t})=V(q_\text{s},q_\text{t})-q_\text{t}p_\text{t}.
	\end{equation}
	Since $q_\text{t}=0$ we conclude that the mixed characteristic is equal to the OPL. The following relationships hold 
	\begin{equation}
		-\frac{\partial W}{\partial q_\text{s}}=p_\text{s}=0, \hspace{1cm} 	-\frac{\partial W}{\partial p_\text{t}}=q_\text{t}=0.
	\end{equation} 
This implies that the OPL $L=V=W$ is a constant. We omit the dependence of $u$  and $w$ on $x$ and $\boldsymbol{\mathit{\hat{t}}}$ for now, and Eq.~(\ref{eq:L}) gives
	\begin{equation}
		\label{eq:d from V}
		d^2=(V-u-w)^2.
	\end{equation}
	
	Since $d$ is the distance between the points $\mathcal{P}_1$ and $\mathcal{P}_2$ on the reflectors, it can be expressed as a function of $u$ and $w,$ i.e., 
	\begin{equation}
		\label{eq: d}
		d^2=(x+wt_1)^2+(-l+u+wt_2)^2.
	\end{equation}
We use Eqs.~(\ref{eq:d from V})-(\ref{eq: d}) to eliminate $d$ and obtain
\begin{equation}
	\label{eq: d eliminated}
	x^2-(V-l)(V+l)+2u(V-l)-2uw+2Vw+2w(xt_1-lt_2)+2uwt_2=0.
\end{equation}
We introduce the reduced OPL denoted by $\beta=V-l$. We divide
Eq.~(\ref{eq: d eliminated}) by $w$ and substite $\widetilde{w}=1/w$, since $w>0$. 
Furthermore, we parametrize the unit direction vector $\boldsymbol{\mathit{\hat{t}}}$ by the stereographic projection $y$ from the south pole \cite[p.~60-62]{ecmi1}. 
The  stereographic projection $y$ and the corresponding inverse projection $\boldsymbol{\mathit{\hat{t}}}$ are given by
\begin{equation}
	\label{eq:stereographic}
y=\frac{t_1}{1+t_2}, \quad	\boldsymbol{\mathit{\hat{t}}}=\begin{pmatrix}t_1\\t_2\end{pmatrix}=\frac{1}{1+y^2}\begin{pmatrix}2y\\1-y^2\end{pmatrix}.
\end{equation}
	Eq.~(\ref{eq: d eliminated}) can now be expressed as
	\begin{equation}
		u\widetilde{w}+\frac{\widetilde{w}}{2\beta}\Big(x^2-\beta(V+l)\Big)+\frac{u}{\beta}\Big(t_2-1\Big)=\frac{1}{\beta}\Big(lt_2-xt_1-V\big).
		\label{eq: u and tilde w with t}
	\end{equation}
Eq.~(\ref{eq: u and tilde w with t}) is factorized by adding the term $(t_2-1)(x^2-\beta(V-l))/ 2 \beta^2$ on both sides of the equation 
\begin{equation} 
	\label{eq: factorized u & w}
	\left(u+\frac{x^2}{2\beta}-\frac{1}{2}(V+l)\right)\left((1+y^2)\left(\widetilde{w}+\frac{t_2-1}{\beta}\right)\right)=-\left(1+\frac{xy}{\beta}\right)^2.
\end{equation}
We introduce the following notation and express	Eq.~(\ref{eq: factorized u & w}) as
	\begin{equation}
	\label{eq: kappa exp}
	\kappa_1(x)\kappa_2(y)=A(x,y).
	\end{equation}
	Clearly, ${A}(x,y)<0$. We investigate whether ${\kappa}_1(x)$ and ${\kappa}_2(y)$ are positive or negative because we want to take the logarithm of these arguments to obtain an equation as in Eq.~(\ref{eq: cost}). From Fig.~\ref{fig:two reflectors} we have $l=wt_2+di_2+u$.
	We also know that 
	\begin{equation}
		\label{eq: beta}
		\beta=V-l=d(1-i_2)+w(1-t_2).	
	\end{equation}
 From Eqs.~(\ref{eq: factorized u & w})-(\ref{eq: beta}), we obtain
	\begin{eqnarray}
		{\kappa}_2(y)&=&(1+y^2)\left(\frac{1}{w}-\frac{1-t_2}{d(1-i_2)+w(1-t_2)}\right)  \nonumber \\
		&=&(1+y^2)\left(\frac{1}{w}-\frac{1}{d\left(\frac{1-i_2}{1-t_2}\right)+w}\right)>0.
	\end{eqnarray}
	Since ${\kappa}_2(y)>0$ and ${A}(x,y)<0,$ ${\kappa}_1(x)<0$, we multiply the negative factors by $-1$ and obtain the relation $-{\kappa}_1(x){\kappa}_2(y)=-{A}(x,y)$. Next, we want to make Eq.~(\ref{eq: kappa exp}) dimensionless. The unit direction vector ${\boldsymbol{\hat{\mathit{t}}}}$ is dimensionless and consequently $y$ is also dimensionless. We scale all other lengths by a factor $\beta$ using $x=\beta\hat{x}$, $V=\beta\hat{V}$, $l=\beta\hat{l}$, $u(x)=\beta\hat{u}(\hat{x})$ and $\widetilde{w}({\boldsymbol{\mathit{\hat{t}}}})=(\hat{w}(y))/\beta$. With Eqs.~(\ref{eq: factorized u & w})-(\ref{eq: kappa exp}), we obtain dimensionless functions, denoted by $\hat{\kappa}_1(\hat{x})$, $\hat{\kappa}_2(y)$, $\hat{A}(\hat{x},y)$ such that $\hat{\kappa}_1(\hat{x})\hat{\kappa}_2(y)=\kappa_1(x)\kappa_2(y)$. Eq.~(\ref{eq: kappa exp}) can be expressed as 
	\begin{subequations}
		\begin{align}
	\label{eq:initial kappa exp}
	-\hat{\kappa}_1(x)\hat{\kappa}_2(y)&=-\hat{A}(x,y),\\
	\intertext{where,} \hat{\kappa}_1(x)&=\frac{u}{\beta}+\frac{x^2}{2\beta^2}-\frac{V+l}{2\beta},\\
	\label{eq:Kappa2}
	\hat{\kappa}_2(y)&=\frac{\beta}{w}\left(1+y^2\right)-2y^2,\\
	\hat{A}(x,y)&=-\Big(1+\frac{xy}{\beta}\Big)^2.
	\end{align}
	\label{eq:kappa exp}
\end{subequations}
We take the logarithm on both sides of Eq.~(\ref{eq:initial kappa exp}) to obtain the desired form in Eq.~(\ref{eq: cost}), and obtain the following relations
\begin{subequations}
	\begin{align}
		\label{eq:u1}
	u_1(x)&=\log(-\hat{\kappa}_1(x)),\\
	\label{eq:u2}
	u_2(y)&=\log(\hat{\kappa}_2(y)),\\
	\label{eq:cost fxn}
	c(x,y)&=\log(-\hat{A}(x,y)).
	\end{align}
\end{subequations}

\subsection{{Energy conservation}} 
Any ray can be parameterized by one  position and one direction coordinate. The position coordinate is the spatial coordinate of the ray on a screen perpendicular to the optical axis. The direction coordinate is the direction cosine of the angle measured counterclockwise that a ray makes with the plane perpendicular to a screen. We introduce the notation $S$ and ${T}$ for the source and target domains, and subscripts $q$ and $p$ corresponding to position and direction domains, respectively. So, we can say that $S_q$$=$$[\chi_{\min},\chi_{\max}]$, $S_p=[\psi_{\min},\psi_{\max}]$, ${T}_q$$=$$  [\zeta_{{\min}},\zeta_{{\max}}]$, and ${T}_p$$=$$[\varphi_{\min},\varphi_{\max}]$. We assume that $\psi=0^\circ\in S_p$ and $\zeta=0\in T_q$. The reason for this will be evident in Eq.~(\ref{eq: rad to f,g}). The source and target position coordinates are denoted by $\chi$ and $\zeta$, respectively, where $\chi\in[\chi_1,\chi_2]\subseteq S_q$ and $\zeta\in[\zeta_{1},\zeta_{2}]\subseteq T_q$. The incident and reflected rays make angles $\psi$ and $\varphi$ with the normals to source and target planes, i.e., the optical axis, respectively, where $\psi\in[\psi_{\min},\psi_{\max}]\subseteq S_p$ and $\varphi\in[\varphi_{\min},\varphi_{\max}]\subseteq T_p$. The radiance $L$ \cite[p.~17]{carmela} is defined as the radiant flux per unit projected length and per unit angle. The radiance at the source and target are denoted by $L_\text{s}(\chi,\psi)$ and $L_\text{t}(\zeta,\varphi)$. Conservation of flux leads to 
	\begin{equation}
		\int_{\psi_{1}}^{\psi_{2}}\int_{\chi_{1}}^{\chi_{2}} L_\text{s}(\chi,\psi)\cos(\psi) \,\text{d}\chi\,\text{d}\psi=	\int_{\varphi_{1}}^{\varphi_{2}}\int_{\zeta_{{1}}}^{\zeta_{{2}}} L_\text{t}(\zeta,\varphi)\cos(\varphi) \,\text{d}\zeta\,\text{d}\varphi,
		\label{eq: cons flux}
	\end{equation}
for arbitrary subsets in the spatial and angular, source and target domains. For a parallel source and point target, we have the following relations
\begin{subequations}
	\begin{align}
	L_\text{s}(\chi,\psi)&=f(\chi)\delta(\psi),\\
	L_\text{t}(\zeta,\varphi)&=g(\varphi)\delta(\zeta),
	\end{align}
\label{eq: rad to f,g}
\end{subequations}\noindent 
where $f(\chi)$ is the exitance at the source, $g(\varphi)$ is the intensity at the target, \linebreak $\delta(\psi)$ and $\delta(\zeta)$ are Dirac delta functions. We have a source parallel to the optical axis, so $\psi=0^\circ$. Substituting Eq.~(\ref{eq: rad to f,g}) in Eq.~(\ref{eq: cons flux}) leads to 
\begin{equation}
	\int_{\chi_{1}}^{\chi_{2}} f(\chi)\,\text{d}\chi=\int_{\varphi_{1}}^{\varphi_{2}} g(\varphi)\cos(\varphi)\,\text{d}\varphi.
	\label{eq: energy cons 2-ref}
\end{equation}
\begin{figure}[H]
	\vspace{-4mm}
	\centering
	\includegraphics[]{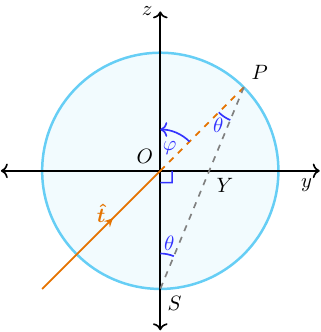}
	\caption{Representation of stereographic projection from the south pole.}
	\label{fig: stereographic projection}
	\vspace{-4mm}
\end{figure} 

We want to rewrite the target intensity $g(\varphi)$ in stereographic coordinates as $\tilde{g}(y)$. First, we express the angle $\varphi$ in terms of the stereographic projection from the south pole. Fig.~\ref{fig: stereographic projection} shows that the stereographic projection is given by $Y(y,0)$. We know that $\bigtriangleup OPS$ is an isosceles
triangle as vectors $\overrightarrow{OS}$ and
$\overrightarrow{OP}$ have a unit length. Therefore, angles $\angle OPY$ and $\angle OSY$ are equal and we denote them by $\theta$. We notice that $y$$=$$\tan(\theta)$ and from $\bigtriangleup OPS$, we observe $\theta$$=$$\frac{\varphi}{2}$. This leads to
\begin{equation}
y=\tan\left(\frac{\varphi}{2}\right).
\label{eq: ster and angle reln}
\end{equation}

We now indicate the spatial source domain with $\mathcal{S}$$=$$[x_{\min},x_{\max}]$ and the stereographic target domain with $\mathcal{T}$$=$$[y_{\min},y_{\max}]$, where $y_{\min}$$=$$\tan(\varphi_{\min}/2)$ and $y_{\max}$$=$$\tan(\varphi_{\max}/2)$. We conserve energy globally (as elaborated in Sect.~\ref{sec: inverse: one reflector}) and use Eq.~(\ref{eq: ster and angle reln}) to make a coordinate transformation in Eq.~(\ref{eq: energy cons 2-ref}), which results in
	\begin{equation}
		\label{eq:energycons2}
		\int_{\mathcal{S}}f(x)\,\text{d}x=\int_{\mathcal{T}}\tilde{g}(y)\frac{2(1-y^2)}{(1+y^2)^2}\,\text{d}y.
	\end{equation}
	Substituting the mapping $y=m(x)$ in Eq.~(\ref{eq:energycons2}) gives
	\begin{equation}
		\label{eq:mappingode2}
		m'(x)=\pm\frac{(1+m^2(x))^2}{2(1-m^2(x))}\frac{f(x)}{\tilde{g}(m(x))},
	\end{equation}
subject to the transport boundary condition $m(\partial \mathcal{S})=\partial \mathcal{T}$, analogous to the single-reflector case in Sect.~\ref{sec: inverse: one reflector}. The numerical solution is \textit{interpolated} to determine the mapping $y=m(x)$.  

\noindent \subsection{{Freeform reflectors}}
Eq.~(\ref{eq: cost}) has many solutions, but we restrict ourselves to a c-convex pair $(u_1,u_2)$ in order to ensure the existence and uniqueness of the optical map \cite[p.~58-64]{nitin}. A c-convex pair is defined as follows: 
\begin{subequations}
\begin{align}
	u_1(x)&=\underset{y\in\mathcal{T}}\max\Big(c(x,y)-u_2(y)\Big), \hspace{1cm} \forall x\in \mathcal{S},	\\ 
	u_2(y)&=\underset{x\in\mathcal{S}}\max\Big(c(x,y)-u_1(x)\Big), \hspace{1cm} \forall y\in \mathcal{T}.
	\label{eq:convexsoln}
\end{align}
\end{subequations}
Eq.~(\ref{eq:convexsoln}) necessarily requires $x$ to be a stationary point of $c(x,y)-u_1(x)$, i.e.,
\begin{equation}
\label{eq:stationary point}
u_1'(x)=\frac{\partial c}{\partial x}(x,y).
\end{equation}
We solve Eq.~(\ref{eq:mappingode2}) numerically and find the mapping $y=m(x)$. Substituting this mapping in Eq.~(\ref{eq:stationary point}) gives us the ODE 
\begin{equation}
	\label{eq: stationary point m(x)}
u_1'(x)=\frac{\partial c}{\partial x}(x,m(x)).
\end{equation}
 We can compute a numerical solution for $u_1(x)$, by solving an IVP if we prescribe an initial condition for Eq.~(\ref{eq: stationary point m(x)}). This choice allows us to set the location of the first reflector. In this paper, we would like to fix the location of the vertex of the first reflector. For a parallel beam with $\boldsymbol{\hat{\mathit{s}}}=(0,1)^{T}$ emitted from a symmetric source domain, the central ray from the point $x_0=0$ hits the vertex. We shift the numerical solution by choosing $u_1(0)$, such that $u(0)=u_0$, implying that the vertex of the first reflector is at a height $u_0$ from the source plane. 

The \textcolor{orange!90!black}{\bf \textit{first reflector} $\mathcal{R}_1$} is given by $\boldsymbol{r}(x)=(x,-l+ u(x))^T$, where 
\begin{equation}
	u(x)= -\beta\exp(u_1(x))-\frac{x^2}{2\beta}+\frac{V+l}{2}.
	\label{eq: reflector1 u(x)}
\end{equation}

We substitute the calculated $u_1(x)$ in Eq.~(\ref{eq: cost}) and obtain $u_2(m(x))$$=$$c(x,m(x))-u_1(x)$. Utilizing Eq.~(\ref{eq:u2}), we obtain a relation for $w(\boldsymbol{\hat{t}})$ dependent on $u_2(m(x))$. The \textcolor{orange!90!black}{\bf \textit{second reflector} $\mathcal{R}_2$} is given by $\boldsymbol{r}(\boldsymbol{\hat{t}})=\left(-w(\boldsymbol{\hat{t}})t_1,-w(\boldsymbol{\hat{t}})t_2\right)^T$, where
\begin{equation}
	w(\boldsymbol{\hat{t}})=\beta\left(1+m(x)^2\right)\left(\exp(u_2(m(x)))+2m(x)^2\right)^{-1}.
\end{equation}

	 \subsection{{Condition for symmetry}}
	Consider a symmetric source domain $x\in[-a,a]$, for some $a>0$. We aim to design rotationally symmetric reflectors, implying that we impose the conditions that the downward and upward unit normal vectors to the first and the second reflector at $x=0$ are $(0,-1)^T$ and $(0,1)^T$, resepctively. Subsequently, using the law of reflection, we obtain $m(0)=0$. We prove that if $\Upsilon(x)=f(x)/(2\tilde{g}(m(x)))$ is an even function, then both reflectors will be symmetric about the optical axis. From Eq.~(\ref{eq:mappingode2}), $\Upsilon(x)$ can be expressed as
\begin{equation}
	 	\Upsilon(x)=\frac{1-m^2(x)}{(1+m^2(x))^2}m'(x)=\frac{\text{d}}{\text{d}x}\frac{m(x)}{1+m^2(x)}.
	 		\label{eq: mapping with ratio 2: rho(x)}
\end{equation}
Similarily, we have
\begin{eqnarray} 
	 		\Upsilon(-x)=\frac{1-m^2(-x)}{(1+m^2(-x))^2}m'(-x)=-\frac{\text{d}}{\text{d}x}\frac{m(-x)}{1+m^2(-x)}.
	 	\label{eq: mapping with ratio 2: rho(-x)}
\end{eqnarray}
Consider $\Upsilon(x)=\Upsilon(-x)$. Using Eqs.~(\ref{eq: mapping with ratio 2: rho(x)})-(\ref{eq: mapping with ratio 2: rho(-x)}) we obtain the following 
\begin{subequations}
	\begin{align}
\frac{\text{d}}{\text{d}x}\frac{m(x)}{1+m^2(x)}&=-\frac{\text{d}}{\text{d}x}\frac{m(-x)}{1+m^2(-x)}	,\\
\intertext{which after integration gives,}
\frac{m(x)}{1+m^2(x)}&=-\frac{m(x)}{1+m^2(x)}+C,
\label{eq: mapping 2 int}
\end{align}
\end{subequations}
where $C$ is a constant of integration. The initial condition on the optical map at $x=0$, implies that $C=0$. Using this, we factorize Eq.~(\ref{eq: mapping 2 int}) and obtain 
\begin{equation}
\Big(m(x)+m(-x)\Big)\Big(1+m(x)m(-x)\Big)=0.
\label{eq: factorize m}
	\end{equation}
Since $y=m(x)$ is the stereographic projection on the unit circle, $m(x)\in(-1,1)$. This implies that $m(x)=-1/m(-x)$ is not possible. Therefore, the optical map is an odd function, i.e., $m(x)=-m(-x)$.

Using Eqs.~(\ref{eq: stationary point m(x)}) and (\ref{eq:u1}), and substituting $y=m(x)$ we obtain
\begin{equation}
	u_1'(x)=\frac{2m(x)}{\beta+xm(x)}.
	\label{eq: u1' exp}
\end{equation}
Since $m(x)$ is an odd function, Eq~(\ref{eq: u1' exp}) implies  $u_1'(x)=-u_1'(-x)$ and subsequently $u_1(x)=u_1(-x)$.
Using this in Eq.~(\ref{eq: reflector1 u(x)}) we conclude that $u(x)=u(-x)$. Therefore, the first reflector $\mathcal{R}_1$ given by $\boldsymbol{r}(x)=(x,-l+ u(x))^T$ is symmetric.

Using Eq.~(\ref{eq:cost fxn}) and the odd function $y=m(x)$, we determine
\begin{equation}
	c(-x,m(-x))=c(x,m(x)).
	\label{eq: sym reln for c}
\end{equation}
With the relations given by Eqs.~(\ref{eq: cost}) and (\ref{eq: sym reln for c}), and odd functions $u_1(x)$ and $m(x)$, we obtain the following for $u_2(m(x))$
\begin{equation}
		\label{eq: u2}
	\begin{aligned}
	u_2(m(-x))&=c(-x,m(-x))-u_1(-x)\\
&=c(x,m(x))-u_1(x)\\
&=u_2(-m(x)).
	      \end{aligned}
\end{equation}
Eq.~(\ref{eq: u2}) implies that $u_2(y)$ is an even function, i.e., $u_2(y)=u_2(-y)$. We know that $u_2(y)$ and $y^2$ are even functions. Therefore, from Eqs.~(\ref{eq:Kappa2}) and (\ref{eq:u2}), we conclude that $w(\boldsymbol{\hat{t}}(y))$ is also an even function. 
With Eq.~(\ref{eq:stereographic}) for stereographic projection of $y$, we have the following relations
\begin{equation}
	t_1(-y)=-t_1(y), \quad t_2(-y)=t_2(y).
		\label{eq: stereographic even/odd}
\end{equation}
Since $w(\boldsymbol{\hat{t}}(y))=w(\boldsymbol{\hat{t}}(-y))$, Eq.~(\ref{eq: stereographic even/odd}) leads to
\begin{equation}
	\begin{aligned}
 \boldsymbol{r}(\boldsymbol{\hat{t}}(-y))&=\left(-w(\boldsymbol{\hat{t}}(-y))t_1(-y),-w(\boldsymbol{\hat{t}}(-y))t_2(-y)\right)^T\\
&=\left(w(\boldsymbol{\hat{t}}(y))t_1(y),-w(\boldsymbol{\hat{t}}(y))t_2(y)\right)^T.
\end{aligned}	\end{equation}
Therefore, the second reflector $\mathcal{R}_2$ is also symmetric. 

\section{Algorithm to minimize aberrations} \label{sec:algorithm}
		We propose an iterative algorithm to find optimal energy distributions to design the freeform reflectors that minimize spot sizes for various sets of off-axis parallel ray bundles. We summarize the algorithm using the flowchart in Fig.~\ref{fig:algorithm} and elaborate further by dividing it into the following steps:
		\begin{figure}[H]
			\centering
			\includegraphics[]{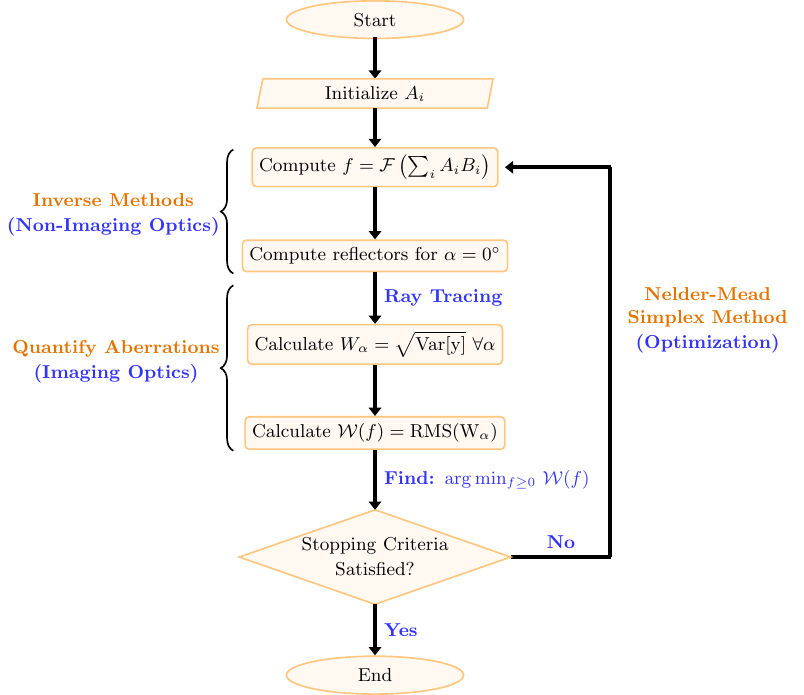}
			\vspace{3mm}
			\caption{Flowchart for the algorithm to minimize aberrations.}
			\label{fig:algorithm}
			\vspace{-4mm}
		\end{figure}
	\begin{enumerate}
		\item \textbf{\textcolor{blue!80}{Compute reflectors for the base case}}\newline
		 In Sect.~\ref{sec: inverse: one reflector}-\ref{sec: inverse: two reflectors}, a detailed mathematical model of two types of optical systems (see Fig.~\ref{fig: both systems}) is given. The systems consist of on-axis parallel rays (base case $\alpha=0^{\circ}$) and give a non-aberrated spot for any given energy distribution. Consider energy distributions $f$ and $g$ at the source and target, respectively. The shape of the reflectors depends the the ratio $\Upsilon$. So, we choose $g$ as a constant and only change $f$. Therefore, the energy distribution at the source is used to compute freeform reflectors. 
		\item	\textbf{\textcolor{blue!80}{RMS spot size}}\\
		An incoming parallel beam under an angle $\alpha$ is directed towards a reflector that has been computed using the energy distribution $f$. The RMS spot size $\left(W_{\alpha}\right)_{{f}},$ corresponding to this beam is estimated by the standard deviation of the ray traced target coordinates $y,$ i.e.,
		\begin{equation}
			\left(W_{\alpha}\right)_{{f}}=\sqrt{\text{Var}[y]}.
		\end{equation}
		\item	\textbf{\textcolor{blue!80}{Merit function}}\\
		The on-axis and off-axis parallel light rays are characterized by angles $\alpha_j$, where $j=1,\mathellipsis,n$. The deviation in aberrations for various ray beams are measured by the merit function 
		\begin{equation}
			\mathcal{W}({f})=\left(\frac{1}{n}\sum_{j=1}^{n}\left(W_{\alpha_j}\right)_{{f}}^2\right)^{1/2}. 
		\end{equation}
		
		\item \textbf{\textcolor{blue!80}{Optimization method}}\\
		In order to obtain the minimum value of the merit function, we solve the optimization problem
		\begin{equation} 
			\text{find}\,\quad \mathrm{arg}\,\underset{f\geq0}\min \,\mathcal{W}({f}). 		
		\end{equation}
The reflectors are computed numerically as a collection of points. We do not have an explicit relation for the shapes of the reflectors and consequently we cannot find an analytical optical map for off-axis rays. The spot sizes for off-axis rays are calculated by ray-tracing through already computed reflectors. In this scenario, a direct dependence of $f$ on spot sizes and the merit function is not known. As a result,  the derivative of $\mathcal{W}$ with respect to $f$ is not available. 
 
 The Nelder-Mead simplex method \cite[p.~502-507]{ecmi3} is employed to solve this optimization problem. This method is a direct search method. It is used for unconstrained problems where the objective function is multi-dimensional and the information about derivatives is either not-known or is computationally expensive to find. For minimizing a function with $m$ variables, a simplex with $m+1$ vertices is constructed. The method moves away from the vertex that gives the highest value of the objective function by eliminating the worst performing vertex of the simplex. In every iteration, a new vertex is introduced by any one of the following decisions, reflection, expansion or contraction of the worst point. The method is terminated if the standard deviation evaluated at each iteration is smaller than some tolerance criterion. The stopping criteria depend on the tolerances for the objective function and the variables, and a limit on the maximum number of iterations. The optimization method was implemented in \textit{Matlab}, and we list the stopping criteria values in Sect.~\ref{sec: results}. 
 
 The Nelder-Mead method performs well for problems with low dimensions. Our problem is computationally quite expensive, since in every iteration we have to compute reflectors and subsequently determine the value of the merit function by ray tracing for several ray beams. So, we try to keep the number of variables for the merit function as small as possible, in a manner that the accuracy is not compromised. 
 
 We consider the source distribution $f$ as an element of the vector space generated by a span of (orthogonal) basis functions. As mentioned earlier, the Nelder-Mead method is used for unconstrained optimization problems. To ensure the constraint ${f}\geq0$, we choose a positive function $\mathcal{F}$,  
		\begin{equation}
			{f}=\mathcal{F}\Big(\sum_{i}A_iB_{i}\Big),
		\end{equation}  
		where $A_i$ are some coefficients and $B_{i}$ are some (orthogonal) basis functions. The method is used to find the optimal energy distribution $f$, by optimizing for coefficients $A_i$. In this paper, we use Legendre polynomials $P_i(x)$ as the basis functions. They are orthogonal with respect to the weighting function $\omega(x)$$=$$1$ in the domain $x\in[-1,1]$.
	\end{enumerate}

	\section{Numerical Results}\label{sec: results}
	We aimed to design reflectors that minimize aberrations for off-axis rays. We tested our method for two optical system configurations, a single-reflector system, and a double-reflector system and compared our results to classical design forms. 
	
 \noindent   \textbf{\textcolor{blue!80}{Input parameters for both optical systems:}} The optimization procedure is dependent on the choice of a positive function $\mathcal{F}$ and orthogonal basis function $B_i$, which consequently determines the energy distributions at source and target. We chose 
     \begin{itemize}
	\item \textcolor{orange!90!black}{\bf Non-uniform source distribution:} $f=\exp(\sum_{i=0}^{4}A_iP_{2i})$.\\  We chose the orthogonal basis functions as even-degree Legendre polynomials $P_{2i}$ to ensure a symmetrical energy distribution and consequently rotationally symmetric reflectors.  
	\item \textcolor{orange!90!black}{\bf Uniform target distribution:} $g$ (or $\tilde{g}$)=constant, chosen such that energy is conserved globally. 
	Target distributions $g$ and $\tilde{g}$ correspond to the single- and double-reflector systems respectively.
	\end{itemize}
We conducted tests for our optimization method with various choices of positive functions $\mathcal{F}$, and orthogonal basis functions. However, we observed consistently similar spot sizes for various parallel beams in optimized designs. Therefore, we do not present results for all choices, since they do not significantly enhance the value of the work presented.
 
A common method to test non-imaging systems is by Monte Carlo ray tracing \cite[p. 33]{carmela}. It requires tracing many rays (typically one million) inside the system to obtain a good accuracy. In imaging optics, a ray trace with precise reflectors and limited number of rays gives an accurate value for the spot size. For our results, the reflectors were computed on a fixed grid with $10^4$ points. We generated off-axis parallel rays from a source of unit length under the angles $\alpha\in\{-1^{\circ},-0.2^{\circ},\mathellipsis,1^{\circ}\}$. Quasi-Monte Carlo ray tracing was used \cite[p. 37-42]{carmela} to calculate the spot sizes for various parallel ray beams. We used $10^3$ rays for computational efficiency.  The optimization method was implemented in \emph{Matlab} using the \emph{fminsearch} routine. The tolerances, \emph{TolFun}, on the value of the merit function and \emph{TolX}, on the value of $A_i$, were $10^{-11}$ and $10^{-6}$, respectively. Only five Legendre polynomials were used in order to reduce the computational effort of the algorithm.

	\subsection{Single-reflector system}\label{ss: results one -reflector}
	For a single-reflector system, parallel on-axis rays incident on a parabolic reflector focus to a point image. A parallel source to near-field target base-case system serves as the best choice for optimizing a single-reflector imaging system to obtain minimum aberrations.
	
	To calculate a freeform reflector using the proposed algorithm, we use the input conditions as follows. Energy conservation was ensured in the source domain $\mathcal{S}=[-1,1]$ and the target domain $\mathcal{T}=[-0.5,0.5]$. The IVPs given in Eqs.~(\ref{eq: mapping ode1})-(\ref{eq: reflector ode}) were solved to calculate the optical map $m(x)$ and freeform reflector $u(x)$. The computations were carried out using \emph{Matlab's} ODE solver \emph{ode45} with $\emph{RelTol}=10^{-13}$ and $\emph{AbsTol}=10^{-14}$. The target plane was chosen as $z=-6$ for the parabolic reflector and the freeform reflector.
	
\begin{figure}[H]
		\centering
		\includegraphics[width=10cm,trim={7cm 0cm 7cm 0cm},clip, height=6.75cm]{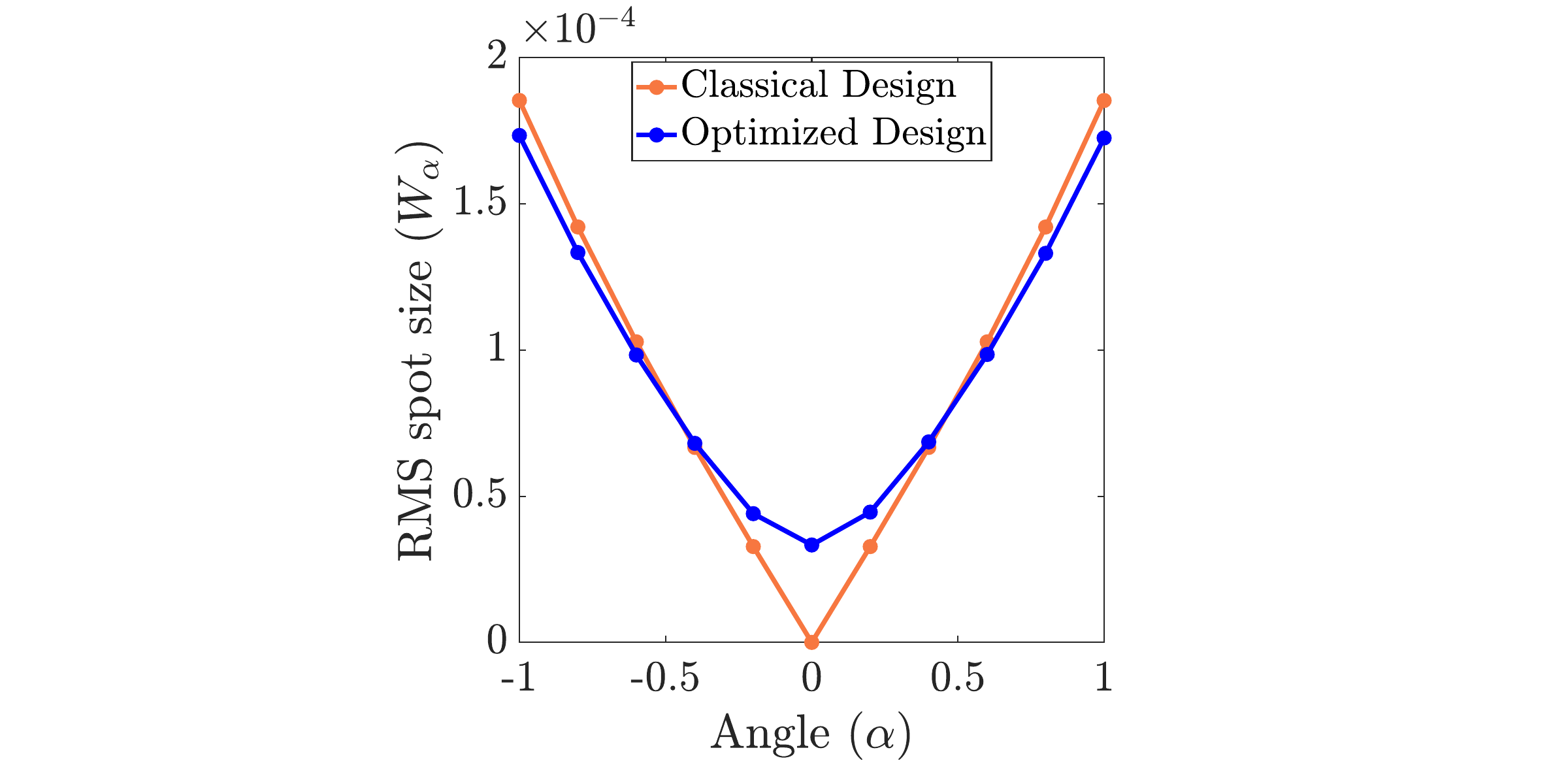}
		\caption{Spot sizes for various angles $\alpha$ for a single-reflector system.}
		\label{result: one reflector}
		\vspace{-4mm}
	\end{figure}

	In Fig~\ref{result: one reflector}, a comparison of the RMS spot size for parallel beams as a function of the angle $\alpha$ is shown. The graph compares the spot sizes for various parallel beams incident on a parabolic reflector (classical design) and an optimized freeform reflector. The design of the freeform reflector is given by our optimization algorithm. From Fig.~\ref{result: one reflector}, we observe that the optimized freeform reflector successfully minimizes the spot size for larger angles. However, it performs poorly for smaller angles when compared to the parabolic reflector. A single-reflector system has limitations for reducing the spot size due to a restricted ability to bend light rays. Therefore, for obtaining a minimum value of the merit function, there is compromised image quality for smaller angles. The values of the merit function for both the design forms is given in Table.~\ref{table:comparison merit function} in Sect.~\ref{ss: summary comparison}. We observe that they do not vary much. This motivates us to use a double-reflector system for minimizing aberrations as it has more flexibility to bend rays and introduces more degrees of freedom for optimization.

	\subsection{Double-reflector system}\label{ss: results two -reflectors}
	In \cite{korsch}, AT is used to design optical systems  that correct third-order aberrations for systems with two or more reflectors.
	These systems consist of reflectors described by a conic section. The shape of a reflector is given by
	\begin{equation}
		z=\mathcal{R}(x)=\frac{x^2}{r+\sqrt{r^2-(1+k)x^2}},
	\end{equation} 
	where $k$ is the deformation constant and $r$ is the radius of curvature.
		\begin{figure}[H]
		\vspace{-3mm}
		\centering
		\includegraphics[trim={0cm 0cm 0cm 1.4cm},clip]{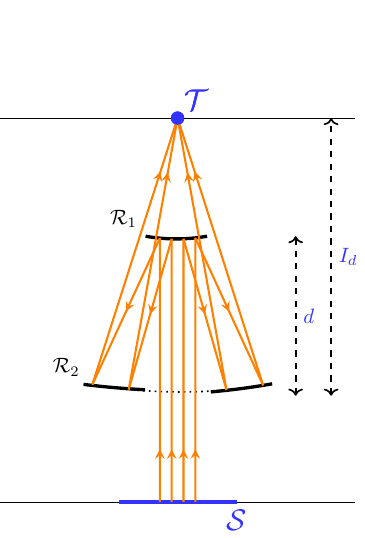}
		\caption{A sketch for the Schwarzschild telescope.}
		\label{telescope}
		\vspace{-4mm}
	\end{figure}
	\begin{table}[h!]
	\vspace{-5mm}
	\renewcommand{\arraystretch}{1.3}
	\setlength{\tabcolsep}{2.5cm}
	\caption{Parameters for a Schwarzschild telescope.}
	\normalsize
	\label{table: telescope parameters}
	\begin{tabular}{|c|}
		\hline
		$\begin{array}{r@{\ } c@{\ } l}
			r_1&=&r_2=-2f\sqrt{2},\\
			d&=&-2f, \\
			I_\text{d}&=&f(1+\sqrt{2}), \\
			k_1&=&(1+\sqrt{2})^2,\\	
			k_2&=&(1+\sqrt{2})^{-2}.
		\end{array}$ 
		\\
		\hline
	\end{tabular}
\end{table}
	 The \textit{Schwarzschild} telescope (see Fig.~\ref{telescope}) is a type of reflecting telescope that uses two curved mirrors to focus light onto a detector. The basic design of this telescope is
	known for an excellent correction of spherical aberration, coma, field curvature, and astigmatism \cite[p.~169]{korsch}. This system is defined completely with the following parameters: vertex radii $(r_1, r_2)$, deformation constants $(k_1,k_2)$, system focal length $(f)$, final image distance from the vertex of the second reflector $(I_\text{d})$, and mirror separation $(d)$. Here, subscripts $1$ and $2$ correspond to the first and second mirror, respectively. For these parameters, the  relations in Table~\ref{table: telescope parameters} hold. We consider $d=4.5$, and the rest of the parameters were calculated using Table~\ref{table: telescope parameters}.
	
 For optimized inverse freeform design using our algorithm, energy conservation was ensured in the source domain $\mathcal{S}=[-0.5,0.5]$ and the stereographic target domain $\mathcal{T}=[-0.25,0.25]$. The even-degree Legendre polynomials used for determining the source distribution $f$ where scaled such that they are orthogonal on $\mathcal{S}=[-0.5,0.5]$. forThe IVPs (Eqs.~(\ref{eq:mappingode2})-(\ref{eq:stationary point})) were solved to calculate the optical map $m(x)$ and $u_1(x)$. The computations were carried out using \emph{Matlab's} ODE solvers \emph{ode45} with $\emph{RelTol}=10^{-12}$ and $\emph{AbsTol}=10^{-16}$. Other parameters affecting the layout of the optical system were chosen as $V$$=$$15$, \linebreak $l$$=$$6$, and $u_0$$=$$5.0680$.
 
\begin{figure}[H]
 	\vspace{-4mm}
 	\centering
 	\includegraphics[width=9.5cm,trim={7cm 0cm 9cm 0.1cm},clip, height=6.4cm]{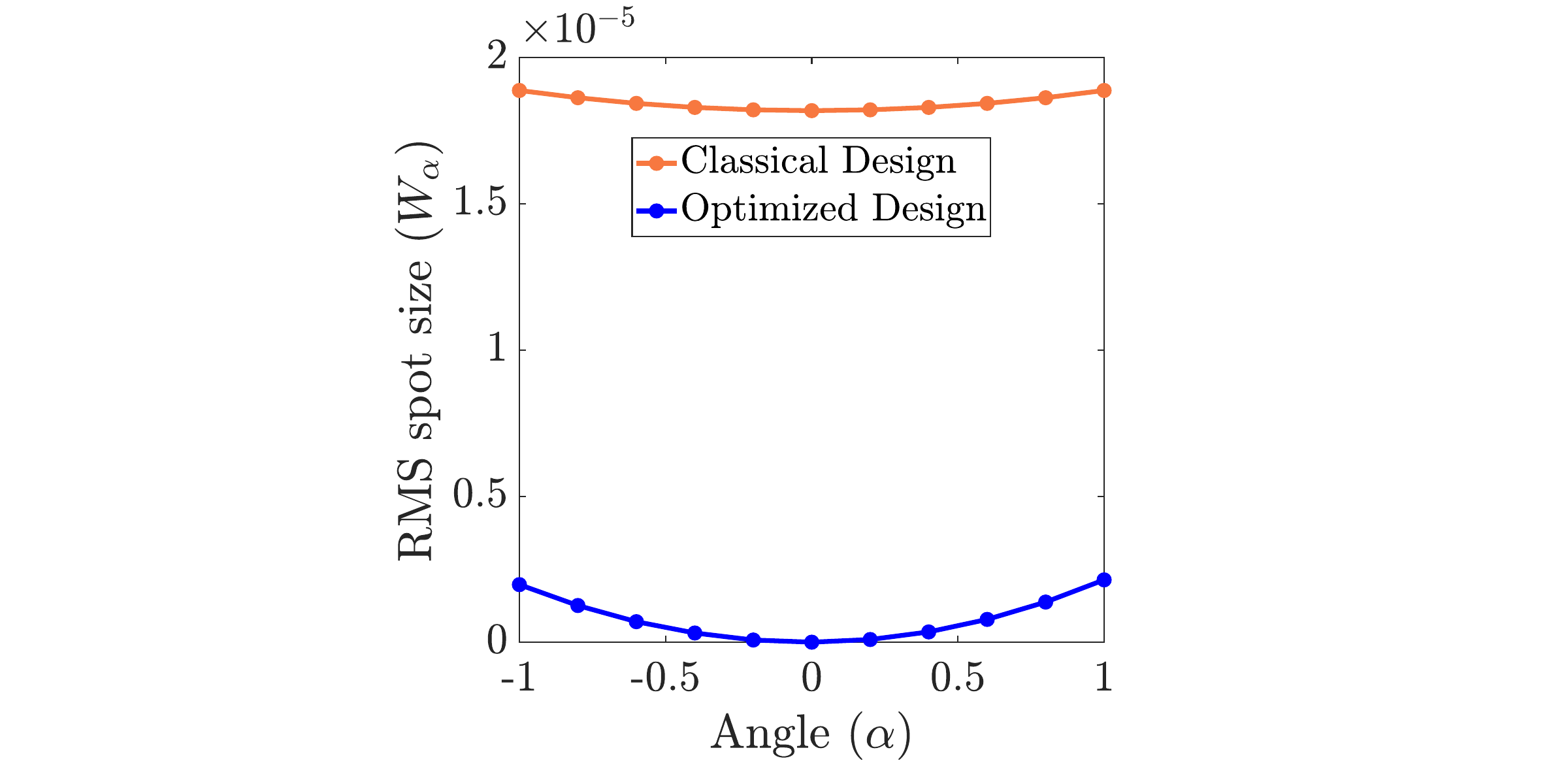}
 	\caption{Spot sizes for various angles $\alpha$ for a double-reflector system.}
 	\label{result: two reflectors}
 	\vspace{-4mm}
 \end{figure}

 We formulate freeform reflectors that give the mininum value of the merit function (see Table.~\ref{table:comparison merit function}). Fig.~\ref{result: two reflectors} shows the spot sizes obtained after ray tracing various parallel beams incident on the optimized freeform reflectors and the classical Schwarzschild telescope.
 
	\subsection{Summary of the comparison}\label{ss: summary comparison}
	
 The proposed algorithm effectively minimizes aberrations for both optical system configurations presented in  Sect.~\ref{ss: results one -reflector}-\ref{ss: results two -reflectors}. Table~\ref{table:comparison merit function} shows the values of the merit function for the classical and the optimized design forms of the single- and the double-reflector imaging systems. We observe that the value of the merit function corresponding to optimized systems is less than the classical design forms. Fig.~\ref{result: one reflector} and \ref{result: two reflectors} depict a comparison of spot sizes corresponding to various angles $\alpha$ for the above-mentioned cases. We conclude that the optimized reflector systems perform better in minimizing aberrations in comparison to classical design forms. A parallel-to-point double-reflector system is better for minimizing aberrations than a parallel-to-near-field single-reflector system.
	\begin{table}[h]
		\vspace{-0.4cm}
		\renewcommand{\arraystretch}{1.3}
		\caption{Value of the merit function $\mathcal{W}(f)$.}
		\label{table:comparison merit function}
		\normalsize
		\begin{tabular}{|c|c|c|}
			\hline
			\textbf{Case} & \textbf{One-reflector} & \textbf{Two-reflectors} \\
			\hline
			\textbf{\textcolor{orange!90!black}{Classical design}} & $1.13346e-04$ & $1.94104e-05$ \\
			\textbf{\textcolor{blue!80}{Optimized design}} & $1.10834e-04$ & $1.10136e-06$\\
			\hline
		\end{tabular}
	\end{table}
	\section{Conclusions and future work}\label{sec: conclusions}
	We proposed an algorithm to optimize two-dimensional reflective imaging systems. We used inverse methods from non-imaging optics to compute fully freeform reflectors. Subsequently, the optical system was optimized using the Nelder-Mead optimization method such that minimum aberrations are produced for off-axis parallel rays beams. We tested our method for two optical system configurations, a parallel-to-near field single-reflector system, and a parallel-to-point double-reflector system. The algorithm minimizes more aberrations in comparison to classical designs for \textit{both} configurations. The double-reflector system is \textit{significantly} better for aberration correction.
	
	Like most other optimization procedures, the Nelder-Mead simplex method has the limitation that it may converge to a local minimum. It is not possible to determine whether a solution given by an optimization procedure is the one corresponding to a global or local minimum. Thus, it could be that the optimized solutions we obtained in Sect.~\ref{sec: results} also correspond to a local minimum. This means that the optimized design forms may not be the best possible designs in terms of the least aberrations or the least value of the merit function. However, we still consider our method as a successful design methodology as the optimized system reduces the merit function by one order of magnitude as compared to the Schwarzschild telescope which has been traditionally well-known for the maximum correction of third-order aberrations.
	
    Future research may involve exploring methods to improve optimization for better designs. It would be interesting to analyze aberrations produced by the optimized reflective systems. This analysis will enable us to introduce design parameters or system layouts that correct certain types of aberrations. Some suggested approaches are changing orientation of the reflectors to off-axis positions or adding more reflectors or lenses in the base case optical systems. An extension of the presented work to three-dimensional optical systems will make the design methodology more applicable.

	\backmatter
	
		\bmhead{Abbreviations}
       2D, two-dimensional; AT, aberration theory; SMS, Simultaneous Multiple Surfaces; GO, geometrical optics; RMS, root-mean-square; ODE, ordinary differential equation; IVP, initial value problem; OPL, optical path length.
	
	\bmhead{Acknowledgments}
	The authors thank Teus Tukker, Ferry Zijp and Koondanibha Mitra for their valuable suggestions.
	
	\section*{Declarations}
	
    	\bmhead{Funding}
	   This work has received funding from Topconsortium voor Kennis en Innovatie (TKI program ``Photolitho MCS" (TKI-HTSM 19.0162)). 
	
		\bmhead{Competing interest}
		The authors declare that they have no competing interests.
		
		\bmhead{Ethics approval} 
		Not applicable.
		 
		\bmhead{Consent to participate}
			Not applicable.
		\bmhead{Consent for publication} Not applicable.
		
		\bmhead{Availability of data and materials}
      	Please contact the corresponding author for data requests.

		\bmhead{Code availability} 	The Matlab scripts for obtaining the results presented in this paper are not publicly available at this time but may be obtained from the corresponding author upon request.
		
		\bmhead{Authors' contributions} All authors contributed equally to this work. All authors read and approved the final manuscript.

\end{document}